# PRACTICAL BIJECTIVE S-BOX DESIGN

[1]Abdurashid Mamadolimov, [2]Herman Isa, [3]Moesfa Soeheila Mohamad
[1,2,3]Information Security Cluster, Malaysian Institute of Microelectronic Systems,
Technology Park Malaysia, 57000, Kuala Lumpur, Malaysia
e-mail: [1]rashid.mdolimov@mimos.my, [2]herman.isa@mimos.my, [3]moesfa@mimos.my

**Abstract.** *Vectorial Boolean functions are usually called Substitution Boxes (S-Boxes) and are used as basic component of block ciphers in Cryptography. The ciphers that are Substitution-permutation networks use bijective S-Boxes i.e., Boolean permutations. S-Boxes with low differential uniformity and high nonlinearity are considered as cryptographically strong. In this paper we study some properties of S-Boxes. We construct $8\times 8$ bijective cryptographically strong S-Boxes. Our construction is based on using non-bijective power functions over the finite field.*

## 1 Introduction

A vectorial Boolean function is a map from $\mathbb{F}_2^n$ to $\mathbb{F}_2^m$, where $\mathbb{F}_2$ is finite field with two elements. Vectorial Boolean functions are usually called S-Boxes and are used as a basic component of block ciphers in Cryptography. The ciphers that are Substitution-Permutation networks use bijective S-Boxes which in effect are Boolean permutations. The $n \times n$ bijective S-Boxes are Boolean permutations on $\mathbb{F}_2^n$.

An important condition on S-Boxes is high resistance to differential and linear cryptanalysis [1], which are the main attacks on block ciphers. The differential cryptanalysis is based on the study of how difference in an input can affect the resultant difference at the output. The functions with low differential uniformity [1]

$$\Delta_F = \max_{a,b \in \mathbb{F}_2^n, a \neq 0} \left| \{ x \in \mathbb{F}_2^n : F(x+a) + F(x) = b \} \right|$$

has good resistance to the differential attack.

The linear cryptanalysis is based on finding affine approximations to the action of a cipher. The functions with high nonlinearity [1]

$$N_F = 2^{n-1} - \frac{1}{2} \max_{a,b \in \mathbb{F}_2^n, b \neq 0} \left| \sum_{x \in \mathbb{F}_2^n} (-1)^{b \cdot F(x) + a \cdot x} \right|$$

possess a good resistance to the linear attack.

S-Boxes with low differential uniformity and high nonlinearity are considered as cryptographically strong. S-Boxes with $8 \times 8$ size can be considered as strong if they have differential uniformity at most 10 and nonlinearity at least 100. Note that known best pair of these parameters is 4, 112 and S-Boxes with differential uniformity below 10 and nonlinearity above 100 are very rare.

Several methods to generate cryptographically strong S-Boxes exist [5-10], such as random generation, the use of finite field operations, as well as heuristic algorithms. The power map $x \to x^d$, where $x \in \mathbb{F}_{2^n}$ has been systematically studied in [1,2]. The power map

$$x^{2^n-2} = \begin{cases} x^{-1}, & \text{if } x \neq 0 \\ 0, & \text{otherwise} \end{cases}, \text{ where } x \in \mathbb{F}_{2^n},$$

which in fact is the inverse function, has the known highest nonlinearity and lowest differential uniformity. Inverse mapping was first studied by L.Carlitz and S.Uchiyama in 1957 and it is unique best bijective S-Box since. An example of such $8 \times 8$ S-Box is Rijndael's S-Box [3], which possess differential uniformity 4 and nonlinearity 112.

In this paper, we find the number of *n*-variable non-affine Boolean permutations up to multiple *n*. We study power and binomial functions in $\mathbb{F}_{2^n}$. There is no bijective power function which could be used as strong S-Box, except inverse function. However, there are non-bijective functions with highest nonlinearity and lowest differential uniformity. We obtain strong bijective S-Boxes using non-bijective power functions. Our $8 \times 8$ S-Boxes have differential uniformity 8, nonlinearity 102 and affinely inequivalent to any sum of a power functions and an affine functions.

In this paper we present the construction of 8x8 S-boxes, however, the results are proven for any size n.

## 2 Preliminaries

### 2.1 Vectorial Boolean functions

Let $F_2$ and $F_{2^n}$ be the finite fields with 2 and $2^n$ elements, respectively. Let $(F_2^n,+)$ be the vector space over $F_2$, where + is used to denote the addition operator over both $F_2$ and the vector space $F_2^n$. An *n*-variable Boolean function (filter) is a map

$$f = f(x_1,...,x_n) : F_2^n \to F_2.$$

The Hamming weight *wt(f)* of a Boolean function *f* on *n* variables is the weight of this string, that is, the size of the support $sp(f) = \{x \in F_2^n : f(x) = 1\}$ of the function. The function *f* is said to be balanced if $wt(f) = 2^{n-1}$. The Hamming distance between two Boolean functions *f* and *g* is $d(f,g) = |\{x : f(x) \neq g(x)\}|$. Clearly, the distance between Boolean functions *f* and *g* is equal to the weight of sum of these functions $d(f,g) = wt(f+g)$.

An (*n,m*) vectorial Boolean function (S-Box) is a map

$$F = F(x_1,...,x_n) = (f_1(x_1,...,x_n),...,f_m(x_1,...,x_n)) : F_2^n \to F_2^m.$$

Clearly, each component function $f_i$, $i = 1,...,m,$ is an *n*-variable Boolean function. An (*n,n*) vectorial Boolean function is called *n*-variable Boolean transformation.

An (*n,m*) vectorial Boolean function is called balanced if it takes every value of $F_2^m$ the same number of times. If a Boolean transformation is balanced then it takes every value of $F_2^n$ one time. A balanced *n*-variable Boolean transformation is called *n*-variable Boolean permutation. Clearly, *n*-variable Boolean permutation is bijective function from $F_2^n$ into itself.

The unique representation of *n*-variable Boolean function *f* as a polynomial over $F_2$ in *n* variables of the form $f(x_1,...,x_n) = \sum_{\alpha \in F_2^n} c(\alpha)(\prod_{i=1}^n x_i^{\alpha_i})$ is called the algebraic normal form (ANF) of *f*. The degree of the ANF of *f* is denoted by $d^\circ(f)$ and is called the algebraic degree of the function *f*.

We call Boolean function *f* as affine if $d^\circ(f) \leq 1$ and linear if it is affine and *f*(0)=0.

An (*n,m*) vectorial Boolean function $F = (f_1,...,f_m)$ is called affine (linear) if each component function $f_1,...,f_m$ is affine (linear). We concentrate on non-affine Boolean permutations.

### 2.2 Differential Uniformity

The differential uniformity (DU) $\Delta_F$ of (*n,m*) vectorial Boolean function *F* is defined as

$$\Delta_F = \max_{a \in F_2^n, a \neq 0, b \in F_2^m} |\{x \in F_2^n : F(x+a) + F(x) = b\}|.$$

For any (*n,m*) vectorial Boolean function *F*, the DU $\Delta_F$ of *F* satisfies

$$\max\{2, 2^{n-m}\} \leq \Delta_F \leq 2^n.$$

If *F* is linear then $\Delta_F = 2^n$. If $n = m$ then $\Delta_F \geq 2$. All known permutations with $\Delta_F = 2$ are defined with odd *n*. It is conjectured that for any Boolean permutation *F* with even *n* $\Delta_F \geq 4$.

### 2.3 Nonlinearity

Let $A(n)$ be the set of all *n*-variable affine Boolean functions. The nonlinearity (NL) $N_f$ of an *n*-variable Boolean function *f* is defined as

$$N_f = \min_{g \in A(n)} d(f,g).$$

In other words, the NL of function *f* is a distance between function *f* and the set *A(n)* of all *n*-variable affine Boolean functions. Clearly, $N_f = 0$ if and only if *f* is an affine function. It is known that for any *n*-variable Boolean function *f*, the NL $N_f$ satisfies the following relation: $N_f \leq 2^{n-1} - 2^{n/2-1}$. Functions achieving the equality are called bent functions which exist when *n* is even. However, bent functions are not balanced. Let *f* be a balanced *n*-variable Boolean function ($n \geq 3$). Then the NL of function *f* is given by

$$N_f \leq \begin{cases} 2^{n-1} - 2^{n/2-1} - 2 & n \text{ even} \\ \lfloor 2^{n-1} - 2^{n/2-1} \rfloor & n \text{ odd,} \end{cases}$$

where $\lfloor x \rfloor$ denotes the largest even integer less than or equal to *x*.

The NL of an (*n,m*) vectorial Boolean function *F*, is defined as $N_F = \min_{c \in F_2^m, c \neq 0} N_{c \cdot F}$.

In other words, the NL of function $F$ is a distance between the set of all non-constant linear combinations of component functions of $F$ and the set $A(n)$ of all $n$-variable affine Boolean functions. This shows that $N_F = 0$ if $F$ is affine. However, the condition $N_F = 0$ does not explain the affinity of $F$. It is known that for any $(n,m)$ vectorial Boolean function $F$, the NL, $N_F$ satisfies $N_F \leq 2^{n-1} - 2^{n/2-1}$. Functions achieving the equality are called perfectly nonlinear and can exist only when $n$ is even and $m \leq \frac{n}{2}$. If $n$ is odd and $n=m$ then we have $N_F \leq 2^{n-1} - 2^{\frac{n-1}{2}}$. Functions with NL $2^{n-1} - 2^{\frac{n}{2}}$ are known for even $n$ and $n=m$, and it is conjectured that this value is the highest possible NL.

## 3 The number of non-affine Boolean Permutations

If we denote the set of all $(n,m)$ vectorial Boolean functions, $n$-variable Boolean transformations and $n$-variable Boolean permutations as $BF(n,m)$, $BT(n)$ and $BP(n)$ respectively, then we have $|BF(n,m)| = 2^{m \cdot 2^n}$, $|BT(n)| = 2^{n \cdot 2^n}$ and $|BP(n)| = 2^n!$.

We denote the set of all non-affine $n$-variable Boolean permutations by $NABP(n)$. Note that $NABP(n) \subset BP(n) \subset BT(n)$.

**Theorem 3.1.** Let $\mu(n) = 2^n! - (2^{n-1}!)^2 \cdot (2^{n+1} - 2)$. Then the number of non-affine Boolean permutations satisfies

$$\mu(n) \leq |NABP(n)| \leq n \cdot \mu(n).$$

Proof: For proving the left side of inequality, it is enough to show that we can construct $\mu(n)$ different non-affine $n$-variable Boolean permutations. Clearly, an $n$-variable Boolean permutation is just permutation of $\mathbb{F}_2^n$ vectors. Our method of construction contains two steps:

(i) Choose balanced non-affine $n$-variable Boolean function as first component function $f_1$ of Boolean permutation.

(ii) Choose two permutations of $\mathbb{F}_2^{n-1}$ vectors and set the permuted vectors as values of $(0, f_2, ..., f_n)$ and $(1, f_2, ..., f_n)$, respectively.

The resulting function $F = (f_1, f_2, ..., f_n)$ is a non-affine Boolean permutation.

Any non-constant affine function is balanced. Since, $|A(n)| = 2^{n+1}$ and the number of constant affine functions is 2, the number of balanced affine Boolean functions is $2^{n+1} - 2$ while the number of $n$-variable balanced Boolean functions is $\binom{2^n}{2^{n-1}}$. Therefore the number of balanced $n$-variable non-affine Boolean function is $\binom{2^n}{2^{n-1}} - (2^{n+1} - 2)$. The number of permutations in step ii) is $(2^{n-1}!)^2$. Thus, we have

$$(2^{n-1}!)^2 \cdot \left( \binom{2^n}{2^{n-1}} - (2^{n+1} - 2) \right) = 2^n! - (2^{n-1}!)^2 \cdot (2^{n+1} - 2) = \mu(n)$$

distinct non-affine Boolean permutations.

To prove the right side of inequality, we first construct $n \cdot \mu(n)$ non-affine Boolean permutations. Then we show that each non-affine Boolean permutation can be obtained by our construction. In the above construction if we take $i$-th component as balanced non-affine fixed function for each $i=1,2,...,n$ then we have $n \cdot \mu(n)$ non-affine Boolean permutations. Let $F = (f_1,...,f_n)$ be any non-affine Boolean permutation. Then $F$ has at least one non-affine component function $f_i$. Clearly, the Boolean permutation $F = (f_1,...,f_n)$ can be obtained by permuting the vectors of $\mathbb{F}_2^n$ such that in the obtained Boolean permutations $i$-th component function is same with $f_i$. □

Table 1 shows the number of functions of three classes for some small $n$.

Table 1
The number of functions in the three classes

| n | $|BT(n)|$ | $|BP(n)|$ | $|NABP(n)|$ |
|---|---|---|---|
| 1 | 4 | 2 | 0 |
| 2 | 256 | 24 | 0 |
| 3 | 16,777,216 | 40,320 | $\geq 32,256$ |
| 8 | $\approx 10^{614}$ | $\approx 10^{513}$ | $\approx 10^{512}$ |

## 4 Some properties of power and binomial functions over the finite field $\mathbb{F}_{2^n}$

Let $\mathbb{F}_{2^n}$ be finite field of $2^n$ elements. We consider $x \to x^d$ power functions, where $x \in \mathbb{F}_{2^n}$ and $d$ is positive integer. We note that power function $x^d$ is bijective if and only if $\gcd(2^n - 1, d) = 1$.

**Theorem 4.1.** Let $x^d$ be power mapping over the multiplicative group $\mathbb{F}_{2^n}^* = \mathbb{F}_{2^n}/\{0\}$. Then it is a $q$-to-one function, where $q = \gcd(2^n - 1, d)$.

Proof: It is known that the group $\mathbb{F}_{2^n}^*$ is cyclic. Let we consider the group $\mathbb{Z}_m$ of residue classes modulo $m$. Since the group $\mathbb{Z}_m$ is also cyclic, if $m = 2^n - 1$ then the multiplicative group $\mathbb{F}_{2^n}^*$ is isomorphic to the additive group $\mathbb{Z}_m$. Therefore it is sufficient to show that the mapping $dx$ over the group $\mathbb{Z}_m$ is a $q$-to-one function, where $q = \gcd(m, d)$. This implies from the following well known theorem of Number Theory: The congruence $dx \equiv r \pmod{m}$ has a solution in $\mathbb{Z}_m$ if and only if $\gcd(m,d)$ divides $r$. In this case, the number of solutions in $\mathbb{Z}_m$ is exactly $\gcd(m,d)$ [4]. □

Consider the binomial function $x^d + x^{2^i}$ over the finite field $\mathbb{F}_{2^n}$.

**Theorem 4.2.** $\mathrm{DU}(x^d + x^{2^i}) = \mathrm{DU}(x^d)$ and $\mathrm{NL}(x^d + x^{2^i}) = \mathrm{NL}(x^d)$.
Proof:
$$\mathrm{DU}(x^d + x^{2^i}) = \max_{a,b \in \mathbb{F}_2^n, a \neq 0} |\{x \in \mathbb{F}_2^n : (x+a)^d + (x+a)^{2^i} + x^d + x^{2^i} = b\}| =$$
$$= \max_{a,b \in \mathbb{F}_2^n, a \neq 0} |\{x \in \mathbb{F}_2^n : (x+a)^d + x^{2^i} + a^{2^i} + x^d + x^{2^i} = b\}| =$$
$$= (\text{since } x^{2^i} \text{ is linear}) = \max_{a,b \in \mathbb{F}_2^n, a \neq 0} |\{x \in \mathbb{F}_2^n : (x+a)^d + x^d + a^{2^i} = b\}| =$$
$$= \max_{a,b \in \mathbb{F}_2^n, a \neq 0} |\{x \in \mathbb{F}_2^n : (x+a)^d + x^d = b\}| = (\text{since } a^{2^i} \text{ is a constant}) = \mathrm{DU}(x^d).$$
$$\mathrm{NL}(x^d + x^{2^i}) = \min_{c \in \mathbb{F}_2^m, c \neq 0} N_{c \cdot (x^d + x^{2^i})} = \min_{c \in \mathbb{F}_2^m, c \neq 0} \min_{g' \in \Lambda(n)} d(c \cdot (x^d + x^{2^i}), g') = \min_{c \in \mathbb{F}_2^m, c \neq 0} \min_{g' \in \Lambda(n)} wt(c \cdot (x^d + x^{2^i}) + g') =$$
$$= \min_{c \in \mathbb{F}_2^m, c \neq 0} \min_{g' \in \Lambda(n)} wt(cx^d + cx^{2^i} + g') = \min_{c \in \mathbb{F}_2^m, c \neq 0} \min_{g \in \Lambda(n)} wt(cx^d + g) = \min_{c \in \mathbb{F}_2^m, c \neq 0} \min_{g \in \Lambda(n)} d(cx^d, g) = \min_{c \in \mathbb{F}_2^m, c \neq 0} N_{cx^d} = \mathrm{NL}(x^d).$$ □

## 5 Construction bijective $8 \times 8$ strong S-Boxes

It is well known some power functions with lowest DU or highest NL [1,2]. For example, power functions $x^d$, where $d = 2^i + 1$ ($d = 2^{2i} - 2^i + 1$) and $\gcd(i, n) = 1$, are called Gold (Kasami) functions and they have lowest DU 2 and if $n$ is odd, then they have highest NL too. If $d = 2^{n-1} - 1$ (inverse function) or $d = \sum_{k=0}^{n/2} 2^{ik}$ and $\gcd(i, n) = 1$, $n \equiv 0 \bmod 4$ then power permutations $x^d$ have the highest known NL.

It is not difficult to compute DU and NL for all power functions over the finite field $\mathbb{F}_{2^8}$. Note that $8 \times 8$ size S-Boxes can be considered as strong if they have DU $\leq 10$ and NL $\geq 100$. The only 8 bijective power functions (permutations) satisfy the conditions are $x^{127}, x^{254}, x^{253}, x^{251}, x^{247}, x^{239}, x^{223}, x^{191}$. They have DU 4 and NL 112. Actually, all these functions are cyclomatic cosets, which means they are equivalent functions. Therefore there is only one strong bijective power function up to equivalence. Rijndael have used this function [3].

We consider strong non-bijective power functions. There are 24 non-bijective power functions with best parameters (DU, NL)=(2,112). Actually, they are from three non-equivalence classes: $\{x^3, x^6, x^{12}, x^{24}, x^{48}, x^{96}, x^{192}, x^{129}\}$, $\{x^9, x^{18}, x^{36}, x^{72}, x^{144}, x^{33}, x^{66}, x^{132}\}$ and $\{x^{39}, x^{78}, x^{156}, x^{57}, x^{114}, x^{228}, x^{201}, x^{147}\}$. We can choose any representative of these functions for next study. For the construction example, we choose $x^3$.

The power function $x^3$ is Gold and Kasami function therefore it has lowest DU 2. The NL of power function $x^3$ is 112. Let $I$ be image of this function, then $|I| = 86$. By Theorem 4.1, the function $x^3$, as mapping $\mathbb{F}_{2^8}^* \to \mathbb{F}_{2^8}^*$, where $\mathbb{F}_{2^8}^* = \mathbb{F}_{2^8} / \{0\}$, is three-to-one function.

We consider $\alpha \cdot x^3 + \beta \cdot x^4$ function, where $\alpha, \beta \in \mathbb{F}_{2^8}$. From Theorem 4.2, DU($\alpha \cdot x^3 + \beta \cdot x^4$)=DU($x^3$)=2 and NL($\alpha \cdot x^3 + \beta \cdot x^4$)=NL($x^3$)=112. Let $J$ denotes the image of $\alpha \cdot x^3 + \beta \cdot x^4$ then $|J| = 192$. Therefore we have been brought nearer to bijective function without changing DU and NL.

Finally, the remaining duplicate elements in the image are replaced by elements from $\mathbb{F}_{2^8}^* \setminus J$ in such a way that the DU and NL are not compromised significantly.

There are $256^2$ functions of the form $\alpha \cdot x^3 + \beta \cdot x^4$, $\alpha, \beta \in \mathbb{F}_{2^8}$. Our best result is (DU, NL)=(8, 102) and only three functions give this result. The coefficients $(\alpha, \beta)$ of these functions are (50, 89), (167, 16) and (218, 71).

## 6 Conclusions

In this paper, we prove some properties of power and binomial functions over the finite field. We find the number of *n*-variable non-affine Boolean permutations up to multiple *n*. Non-affine Boolean permutations are not rare. However, non-affinity property is not sufficient for strong S-Boxes.

We propose a simple scheme which produces a new cryptographically strong bijective S-Boxes. Construction is based on using non-bijective strong power functions over the finite field.

The resulting S-boxes have DU 8 and NL 102 which are cryptographically strong for use in block ciphers.

## Acknowledgment

We would like to thank Prof. Dr. Mohamed Ridza Wahiddin (MIMOS Bhd) for general guidance. We also thank Dr. Isamiddin Rakhimov (UPM) for useful comments.